\documentclass[prd,english,preprintnumbers,nofootinbib,twocolumn,superscriptaddress]{revtex4-1}

\pdfoutput=1

\usepackage[latin1]{inputenc}
\usepackage{graphicx}
\usepackage{bbm}
\usepackage{tabularx}
\usepackage{hyperref}
\hypersetup{colorlinks=false}

\usepackage{color}

\usepackage{dsfont}

\def\0#1#2{\frac{#1}{#2}}

\def\s0#1#2{\mbox{\small{$ \frac{#1}{#2} $}}}

\newcommand{\m}[1]{\ensuremath{\mathrm{#1}}}

\newcommand{\MSbar}{\overline{\mbox{MS}}}

\usepackage{babel}
\makeatother
\begin{document}

\title{Tensor $O(N)$ model near six dimensions: fixed points and conformal windows from four loops}

\author{John A. Gracey}
\affiliation{Theoretical Physics Division, Department of Mathematical Sciences, University of Liverpool, P. O. Box 147, Liverpool, L69 3BX, United Kingdom}
\author{Igor F. Herbut}
\affiliation{Department of Physics, Simon Fraser University, Burnaby, British Columbia, V5A 1S6, Canada}
\author{Dietrich Roscher}
\affiliation{Institute for Theoretical Physics, University of Cologne, 50937 Cologne, Germany}

\begin{abstract}
In search of non-trivial field theories in high dimensions, we study further the tensor representation of the $O(N)$-symmetric $\phi^4$ field theory introduced by Herbut and Janssen (Phys. Rev. D. 93, 085005 (2016)), by using four-loop perturbation theory in two cubic interaction coupling constants near six dimensions. For infinitesimal values of the parameter $\epsilon=(6-d)/2$ we find infrared-stable fixed point with two relevant quadratic operators for $N$ within the conformal windows $1<N<2.653$ and $2.999<N<4$, and compute critical exponents at this fixed point to the order $\epsilon^4$. Taking the four-loop beta-functions at their face value we determine the higher-order corrections to the edges of the above conformal windows at finite $\epsilon$, to find both intervals to shrink to zero above  $\epsilon\approx 0.15$. The disappearance of the conformal windows with the increase of $\epsilon$ is due to the collision of the Wilson-Fisher $\mathcal{O}(\epsilon)$ infrared fixed point with the $\mathcal{O}(1)$ mixed-stable fixed point that appears at two and persists at higher loops. The latter may be understood as a Banks-Zaks type fixed point that becomes weakly coupled near the right edge of either conformal window. The consequences and issues raised by such an evolution of the flow with dimension are discussed. It is also shown both within the perturbation theory and exactly that the tensor representation at $N=3$ and right at the $\mathcal{O}(\epsilon)$ infrared-stable fixed point exhibits an emergent $U(3)$ symmetry. A role of this enlarged symmetry in possible protection of the infrared fixed point at $N=3$ is noted.
\end{abstract}

\maketitle

\section{Introduction}

The question of the existence of interacting conformally invariant field theories in space-time dimensions $d\geq 4$ has been long standing. Using the random walk representation, it has been, for example, rigorously proven that repulsive $O(N)$ $\phi^4$ field theories are trivial for $N=1$ \cite{aizenman} and $N=2$, \cite{frohlich} i. e. that the only fixed point of the scale transformation in such a field theory is the Gaussian one. On the other hand, there is a bicritical fixed point for the $O(N)$ $\phi^4$ theory at $d>4$ and any $N$, but at a negative (attractive) self-interaction. One may then ask if there is a way to represent this fixed point of the $\phi^4$ theory as an infrared (IR)-stable fixed point of another field theory with the same symmetry, which would serve as its ultraviolet (UV) completion. Indeed, there are at least two instances where this is possible: 1) the $\phi^4$ theory at the IR Wilson-Fisher fixed point is believed to be equivalent to the non-linear sigma model at the UV fixed point, 2) the Gross-Neveu model at the UV fixed point is equivalent to the Gross-Neveu-Yukawa field theory at its IR critical point. \cite{zinn-justin}

For the $O(N)$-invariant self-attractive $\phi^4$ theory this question was raised by Fei, Giombi, and Klebanov \cite{fei}, who Hubbard-Stratonovich decoupled the interaction term, to eventually replace the original theory with
\begin{equation}
\label{BaseModel}
\mathcal{L} = \frac{1}{2}(\partial_\mu z_a)^2 + \frac{1}{2}(\partial_\mu\phi_i)^2 + \frac{g_1}{2} z_a\phi_i\Lambda^a_{ij}\phi_j + \frac{g_2}{6} d^{abc}
z_a z_b z_c
\end{equation}
with  the indices $i,j = 1,2,...N$, the index $a=0$, the matrix $\Lambda^0 _{ij} = \delta_{ij}$, and the symbol $d^{000} =1$. $\phi_i$ are the original $N$ real fields that transform as a vector, and $z_0$ is the real Hubbard-Stratonovich field, scalar under $O(N)$. The $O(N)$ symmetry also allows for two quadratic terms, $m_z ^2 z_a z_a $ and $m_\phi ^2 \phi_i \phi_i$ which are IR-relevant, and which in (1) have been and hereafter will be tuned to zero. To understand the origin of the field theory (1) notice that when $g_2 =0$ and $m_z \neq 0$ the field $z_0$ can be integrated out, to recover the original interacting term $(\phi_i \phi_i)^2$ to the leading order in gradient expansion. The interaction $g_2$  and the kinetic energy term for $z_a$ can be thought of, in turn, as being generated by integration over high-energy modes of the fields $\phi_i$, in close analogy to the passage from the Gross-Neveu to the Gross-Neveu-Yukawa field theories, for example. The same theory was also considered as a toy-model of hadron dynamics by Ma. \cite{ma} We will call it the ``scalar representation".

What is gained by this reformulation of the $\phi^4$ theory is a new possibility for a systematic search for non-trivial fixed points using $(6-d)$-expansion, since both cubic interaction couplings $g_1$ and $g_2$ become marginal in the same dimension  $d=6$. It was found that for large $N$ one-loop beta-functions for the two interaction coupling constants do have a Wilson-Fisher $\mathcal{O}(6-d)$ IR-stable fixed point. (For the purposes of this paper an IR-stable fixed point will be defined as a fixed point with only two relevant couplings, which are the above two masses $m_z$ and $m_\phi$.) As $N\rightarrow 1038.266$, however, the Wilson-Fisher fixed point collides  with another $\mathcal{O}(6-d)$ fixed point with mixed stability, which has one additional relevant direction in the IR. Both fixed points then become complex when $N < 1038.266$. Higher-loop computations \cite{giombi, jag1} conformed to the same scenario, but significantly reduced the value of the critical number of components $N$ above which the real IR-stable fixed point exists at finite $d<6$. Nevertheless, the question of the existence of the IR-stable fixed point of the theory (1) in scalar representation for single-digit values of $N$ remained.

It was further realized by Herbut and Janssen \cite{jag13} that it is equally possible to decouple the interaction term of the $\phi^4$ theory in the tensor  channel, so that one ends up with the theory in the form of (1), but with the index $a=1, ... M_N$ with $M_N = (N-1) (N+2) /2$ as the number of components of the irreducible second-rank tensor under $O(N)$, and the $M_N$ matrices $\Lambda^a $  as a basis in the space of traceless, real, symmetric, $N$-dimensional matrices. The real fields $z_a$ transform then under $O(N)$ as components of a second-rank tensor, and the symbol
\begin{equation}
d^{abc} ~=~ \m{Tr} \left( \Lambda^a \Lambda^b \Lambda^c \right).
\end{equation}
In particular, for $N=2$ the two $\Lambda^a$ matrices are nothing but the two real Pauli matrices, and for $N=3$ the five $\Lambda^a$ matrices are the familiar real Gell-Mann matrices.\cite{janssen} We will call this realization of the $O(N)$ symmetry the ``tensor representation". The combination of the scalar and the tensor  representations of the $O(N)$ model was also considered and studied near six dimensions. \cite{osborn, simms} The pure tensor representation of the $O(N)$ model defined as above will be the main subject of the present study.

One-loop renormalization group (RG) analysis of the tensor representation \cite{jag13} revealed, quite surprisingly and in contrast to the scalar representation \cite{fei}, that the IR-stable fixed point exists in the intervals $1<N<2.653$ and $2.999<N<4$, which include physically interesting low values of $N=2$ and $N=3$. This suggests that there could be examples of $O(N)$-symmetric field theories, albeit in higher (second-rank tensor) representation, that are non-trivial in say five dimensions. This conclusion would follow, of course, only if the obtained IR-stable fixed point at a given $N$ within the above conformal windows survives the increase of the parameter $6-d$, from an infinitesimal to a physical integer value of one or two. This however is not guaranteed. To examine this issue Roscher and Herbut \cite{roscher} have performed perturbative two-loop and functional renormalization group calculations for $N=2$, which is particularly simple because the second cubic term identically vanishes, i. e. $d^{abc} \equiv 0$. They found that the second-loop terms introduce an additional non-trivial mixed-stable fixed point of the beta-function, which is $\mathcal{O}(1)$, and which collides with the $\mathcal{O}(\epsilon)$ IR-stable fixed point as the parameter $\epsilon=(6-d)/2$ is increased. At $\epsilon > \epsilon_c$, with $\epsilon_c \ll 1$, this way the single beta-function at $N=2$ for the coupling $g_1$ had no real zeros left. If correct beyond the particular approximations which were employed this result would conform to one's expectation of triviality of the $O(N)$ theories in $d=5$ or $d=4$, but only at the expense of  having a new mixed-stable, in this case actually UV-stable  finite-coupling fixed point already at the upper critical dimension of $d=6$. In other words, the triviality of the tensor representation of the $O(2)$ model in $d=5$ this way is obtained at the price of non-triviality in $d=6$.

In this work we further study the theory (1) in tensor representation, at general $N$, and by using state-of-the-art perturbation theory in the couplings  $g_1$ and $g_2$ to four-loops. This first allows us to compute Taylor expansions for the critical exponents at the IR-stable fixed point to the order $\epsilon^4$. Setting $N=2$ we recover the previous two-loop result \cite{roscher}, and for all $1< N < 2.653$ we find the same scenario of fixed-point collision and annihilation determining the critical line $N_c (\epsilon)$. A good fit to the numerically computed edge of this conformal window to three loops for example is
\begin{equation}
N_c (\epsilon)= 2.65 - 4 \epsilon^{1/2} +0.75 \epsilon + \mathcal{O}( \epsilon^{3/2} ).
\end{equation}
The mixed-stable fixed point that annihilated the Wilson-Fisher fixed point, while inevitably becoming $\mathcal{O}(1)$ right at $N=2$, is actually weakly coupled near to and left of the right edge of the conformal window at $N=2.653$ (Figure 1). It is another example of a Banks-Zaks fixed point \cite{banks}, and the above expansion may be understood as an extrapolation out of the strictly perturbative regime near the end point $\epsilon=0$ and $N=2.653$, where the line $N_c (\epsilon)$ is determined by the collision of two weakly-coupled fixed points: one being $\mathcal{O}(\epsilon)$ (IR-stable Wilson-Fisher) and the other being $\mathcal{O}(2.653-N)$ (mixed-stable Banks-Zaks).

\begin{figure}[t]
\centering
\includegraphics[width=8.5cm]{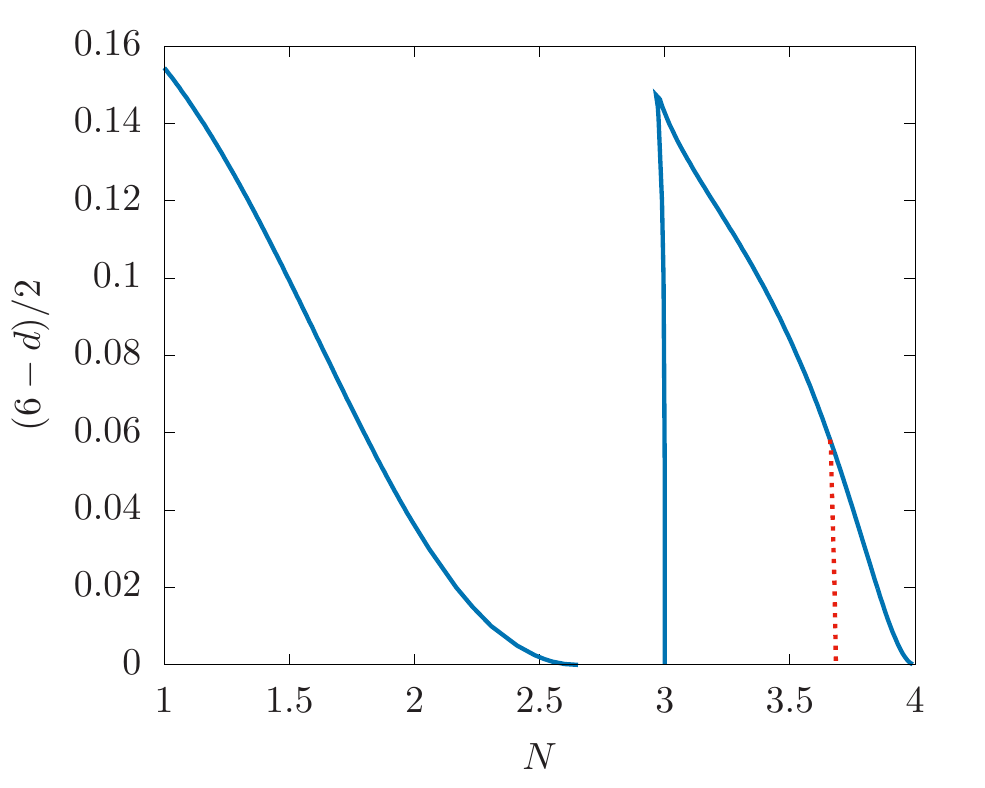}
\caption{Conformal regions (below the full lines) at which the theory (1) in tensor representation for $N>1$ has an IR-stable fixed point, computed using two-loop beta-functions for $g_1$ and $g_2$. At the boundaries of both regions the IR-stable fixed point collides  with another fixed point and becomes complex. At the nearly vertical dashed line which terminates at  $N=3.684$ and $\epsilon=(6-d)/2=0$ the IR-stable and mixed-stable fixed points exchange stability. (For further discussion see the text.)
}
\label{ConfWind}
\end{figure}

The second conformal window for $2.999<N<4$ in Figure 1 also reduces to zero width with the increase of $\epsilon$, but in a more elaborate way, as there are several exchanges of stability between the existing fixed points before the IR-stable fixed point finally disappears above certain value of $\epsilon$. These intricacies  notwithstanding, the final disappearance, or rather complexification of the IR fixed point, is again due to a collision with an $ \mathcal{O}(1) $  fixed point of mixed stability introduced by higher order terms. The physically relevant case of $N=3$ is particularly involved. We first find that at $N=3$ both the vector components $\phi_i$ and the tensor components $z_a$ acquire exactly the same anomalous dimensions, up to the computed fourth order in $\epsilon$. This indicates an emergence of a larger symmetry between these two representations of $O(3)$, which we then show to be $U(3)$:  the fixed point values of the couplings $g_1$ and $g_2$ are precisely such that the two cubic interaction terms in the theory (1) at the fixed point can be rewritten compactly as a single trace of the third power of a traceless, Hermitean, three-dimensional matrix (Eq. 25). The IR-stable $\mathcal{O}(\epsilon)$ fixed point at $N=3$ possesses therefore a larger emergent $U(3)$ symmetry, and the critical exponents reduce to those already computed in \cite{jag1} for the matrix $U(N)$ model. Increasing the value of $\epsilon$, however, we find this fixed point first to exchange stability with another near-by $\mathcal{O}(\epsilon)$ fixed point with mixed stability in the $g_1 - g_2$ plane, before that other fixed point, now IR-stable, collides with an $\mathcal{O}(1)$ fixed point at some critical value of $\epsilon$ and annihilates. Again, near the right corner of the conformal window we find that the collision that determines the boundary is between the Wilson-Fisher $\mathcal{O}(\epsilon)$ fixed point and the $\mathcal{O}(4-N)$ Banks-Zaks-like fixed point. This perturbative region, however, is not smoothly connected to the region that contains $N=3$, since there is an exchange of stability between a pair of fixed points around $N=3.684$. Nevertheless, the mixed-stable fixed point that annihilates the Wilson-Fisher $\mathcal{O}(\epsilon)$ fixed point for $2.999<N<3.684$ is also a Banks-Zaks, albeit weakly coupled at a slightly larger value of $N$ of $4.0057$.

The paper is further organized as follows. In the next section we describe the four-loop renormalization group calculation and give the values of the critical exponents, with the emphasis on $N=3$. In section III we demonstrate the emergence of the $U(3)$ symmetry at the IR-stable $\mathcal{O}(\epsilon)$ fixed point. In section IV we discuss the evolution of conformal windows with parameters $\epsilon$ and $N$, and show how it leads to finite conformal regions in the $N-\epsilon$ plane. We summarize and comment further our findings in the concluding section. Full four-loop expressions for the beta-functions and the exponents at a general $N$ are given in the appendices.

\section{Renormalization}

We turn now to the technical details of the evaluation of the renormalization
group functions for (\ref{BaseModel}). The procedure we followed was similar to
that used for the scalar channel of the $O(N)$-symmetric case at four loops
given in \cite{jag1} and summarize it in this context. As there are two
couplings and fields there are a sizeable number of Feynman diagrams to
evaluate for the $2$- and $3$-point functions which could only be handled by an
automatic computation. All the graphs to four loops for the two $2$-point
functions are generated using the {\sc Fortran} based package {\sc Qgraf},
\cite{jag2}, which is then mapped into the syntax of the symbolic manipulation
language {\sc Form}, \cite{jag3,jag4}, which is the working language for the
automatic evaluation. At this stage $O(N)$ group indices are added and the
group relation
\begin{equation}
\label{groupeval}
\Lambda^a _{ij} \Lambda^a _{kl}= \delta_{ik} \delta_{jl} +   \delta_{il} \delta_{jk} - \frac{2}{N} \delta_{ij} \delta_{kl}
\end{equation}
is implemented within a {\sc Form} module. The
divergent part of each individual graph in dimensional regularization in
$d$~$=$~$6$~$-$~$2 \epsilon$ dimensions is determined by using the Laporta
algorithm, \cite{jag5}. We used both versions of the {\sc Reduze}
implementation, \cite{jag6,jag7}, of the algorithm for the present computation.
In general this systematically uses integration by parts to reduce all Feynman
integrals of a Green's function down to a small base or master set of
integrals. The values of these are determined directly by non-integration by
parts methods. For (\ref{BaseModel}) the set of four loop master integrals for
six dimensional $2$-point functions were given in \cite{jag1} having been
constructed using the Tarasov method, \cite{jag8,jag9}, from the corresponding
four dimensional masters given in \cite{jag10}. In terms of graphs computed for
the $\phi_i$ $2$-point function we evaluated $1$, $5$, $48$ and $637$ graphs at
the respective loop orders from one to four. For the $z_a$ field the
corresponding numbers were $2$, $7$, $60$ and $723$.

This summarizes the process for the wave function renormalization. Clearly
there is a large number of graphs to compute but this is substantially smaller
than the respective number for the coupling constant renormalization. However,
to determine the two beta-functions to four loops we exploited a novel
feature of renormalizing this scalar field theory in six dimensions. As there
are no $4$-point and higher vertices in (\ref{BaseModel}) all the graphs
contributing to the $3$-point functions can be simply generated by making an
insertion on each propagator in the $2$-point function graphs, \cite{jag1}. In
parallel to this the divergent part of each vertex function of
(\ref{BaseModel}) in dimensional regularization can be extracted by nullifying
the external momentum of one of the legs. As the critical dimension of
(\ref{BaseModel}) is six then a propagator $1/(k^2)^2$ in a $3$-point function,
where $k$ is a loop momentum, does {\em not} introduce any infrared
divergences unlike the case if the critical dimension was four. This means
that we do not have to separately generate a substantial number of graphs
using {\sc Qgraf} for the vertex functions. Instead in the computation of the
$2$-point functions we merely make the replacements, \cite{jag1},
\begin{eqnarray}
\label{ccrepl}
\frac{\delta_{ij}}{k^2} & \rightarrow & \frac{\delta_{ij}}{k^2} ~+~
\frac{g_1 \Lambda^{a_e}_{ij}}{(k^2)^2} \nonumber \\
\frac{\Lambda^a_{ik} \Lambda^b_{kj}}{k^2} & \rightarrow &
\frac{\Lambda^a_{ik} \Lambda^b_{kj}}{k^2} ~+~
\frac{g_2}{(k^2)^2} \Lambda^a_{ik} \Lambda^{b_e}_{kl} \Lambda^b_{lj}
\end{eqnarray}
respectively for the $\phi_i$ and $z_a$ propagators where $a_e$ and $b_e$ are
the indices of the external leg. In the second equation the $\Lambda$ matrices
correspond to the matrices from the vertices at either end of the propagator.
They are included so that internally one can distinguish which vertex is being inserted in the {\sc Form} identification. This algebraic expansion is
straightforward to implement within the program in the $2$-point function. It
is truncated in such a way that no more than one term deriving from the
$1/(k^2)^2$ correction is retained. This ensures that only the graphs
corresponding to the vertex functions themselves are selected. Both
replacements are implemented in each $2$-point function with the $g_1$ coupling
constant being deduced from the $\phi_i$ $2$-point function. The other coupling
is renormalized via the $z_a$ $2$-point function. The advantage of proceeding
in this fashion, aside from not having to have a separate vertex function
computation, is that the Laporta algorithm can be applied to both wave function
and coupling constant renormalization in the {\em same} full automatic
computation.

Moreover as we are interested in the renormalization of the masses of the two
fields we can simply implement a replacement similar to (\ref{ccrepl}) for this
which is
\begin{eqnarray}
\label{massrepl}
\frac{1}{k^2} & \rightarrow & \frac{1}{k^2} ~+~
\frac{\mu_\phi^2}{(k^2)^2} \nonumber \\
\frac{1}{k^2} & \rightarrow & \frac{1}{k^2} ~+~ \frac{\mu_z^2}{(k^2)^2}
\end{eqnarray}
for the $\phi_i$ and $z_a$ propagators respectively, \cite{jag1}. The
replacement is truncated at $O(\mu_i^2)$ for $i$~$=$~$\phi$ or $z$. Here the
quantities $\mu_\phi^2$ and $\mu_z^2$ are not masses but merely symbols which
tag the origin of the additional term. These are necessary in order to extract
the renormalization constants of the $2$~$\times$~$2$ mass mixing matrix of
(\ref{BaseModel}). A similar procedure was used in \cite{jag1}. The final part
of the automatic evaluation of the $n$-point functions is the extraction of
the renormalization constants. This is also achieved automatically by the
method discussed in \cite{jag11}. For each of the Green's functions leading to
the wave function, coupling constant and mass renormalization one uses the
Lagrangian with bare parameters such as $g_i$ and the masses. The respective
counterterms are introduced by replacing the bare ones with the corresponding
renormalized variables. The overall divergence remaining at each loop order
after the lower loop counterterm values have been determined is absorbed into
the respective counterterm for that particular $n$-point function. As we have
implemented a zero momentum insertion for the coupling constant renormalization
we determine the renormalization group functions in the $\MSbar$ scheme.

Given the number of Feynman graphs we had to compute for (\ref{BaseModel})
together with the presence of an $O(N)$ symmetry the full expression for each
of the renormalization group functions is cumbersome and are given in the Appendix, and provided
electronically in an attached Supplementary Material file, \cite{jag12}. To
give a flavor of this we record the expressions for the specific case of
$N$~$=$~$3$. First, the two anomalous dimensions are
\begin{widetext}
\begin{eqnarray}
\gamma_\phi(g_i) &=&
 \frac{5}{9} g_1^2
+ 35 \left[
- 17 g_1^2 + 24 g_1 g_2
- 11 g_2^2 \right] \frac{g_1^2}{972}
\nonumber \\
&&
- 5 \left[ 29808 \zeta_3 g_1^4
- 66638 g_1^4
+ 31542 g_1^3 g_2
- 9072 \zeta_3 g_1^2 g_2^2
- 12383 g_1^2 g_2^2
+ 15288 g_1 g_2^3
- 13587 g_2^4 \right] \frac{g_1^2}{52488}
\nonumber \\
&&
+ 5 \left[ 216635472 \zeta_3 g_1^6
+ 12060576 \zeta_4 g_1^6
- 479623680 \zeta_5 g_1^6
+ 185271707 g_1^6
- 192961440 \zeta_3 g_1^5 g_2
\right. \nonumber \\
&& \left. ~~~~~
+ 47029248 \zeta_4 g_1^5 g_2
+ 333931878 g_1^5 g_2
+ 152962992 \zeta_3 g_1^4 g_2^2
+ 84832272 \zeta_4 g_1^4 g_2^2
- 209018880 \zeta_5 g_1^4 g_2^2
\right. \nonumber \\
&& \left. ~~~~~
- 106561007 g_1^4 g_2^2
+ 166779648 \zeta_3 g_1^3 g_2^3
+ 82791072 \zeta_4 g_1^3 g_2^3
- 179951100 g_1^3 g_2^3
+ 20112624 \zeta_3 g_1^2 g_2^4
\right. \nonumber \\
&& \left. ~~~~~
+ 34128864 \zeta_4 g_1^2 g_2^4
- 300464640 \zeta_5 g_1^2 g_2^4
+ 238593145 g_1^2 g_2^4
- 37612512 \zeta_3 g_1 g_2^5
- 20085408 \zeta_4 g_1 g_2^5
\right. \nonumber \\
&& \left. ~~~~~
+ 108293094 g_1 g_2^5
+ 12183696 \zeta_3 g_2^6
+ 3347568 \zeta_4 g_2^6
- 68577061 g_2^6 \right] \frac{g_1^2}{34012224} ~+~ \mathcal{O}(g_i^{10})
\end{eqnarray}
and
\begin{eqnarray}
\gamma_z(g_i) &=&
\left[
 3 g_1^2
+ 7 g_2^2 \right]
\frac{1}{18}
+ 7 \left[
- 42 g_1^4
+ 72 g_1^3 g_2
- 33 g_1^2 g_2^2
- 113 g_2^4 \right] \frac{1}{972}
\nonumber \\
&&
- \left[ 178848 \zeta_3 g_1^6
- 473475 g_1^6
+ 404712 g_1^5 g_2
- 136080 \zeta_3 g_1^4 g_2^2
- 130473 g_1^4 g_2^2
+ 386568 g_1^3 g_2^3
- 176001 g_1^2 g_2^4
\right. \nonumber \\
&& \left. ~~~
+ 371952 \zeta_3 g_2^6
- 1217531 g_2^6 \right] \frac{1}{209952}
\nonumber \\
&&
+ \left[ 119362896 \zeta_3 g_1^8
+ 6030288 \zeta_4 g_1^8
- 239811840 \zeta_5 g_1^8
+ 60083890 g_1^8
- 175624848 \zeta_3 g_1^7 g_2
+ 1469664 \zeta_4 g_1^7 g_2
\right. \nonumber \\
&& \left. ~~~
+ 325111584 g_1^7 g_2
+ 135218160 \zeta_3 g_1^6 g_2^2
+ 99610560 \zeta_4 g_1^6 g_2^2
- 209018880 \zeta_5 g_1^6 g_2^2
- 97863857 g_1^6 g_2^2
\right. \nonumber \\
&& \left. ~~~
+ 320005728 \zeta_3 g_1^5 g_2^3
+ 120022560 \zeta_4 g_1^5 g_2^3
- 327324144 g_1^5 g_2^3
+ 105561792 \zeta_3 g_1^4 g_2^4
+ 55275696 \zeta_4 g_1^4 g_2^4
\right. \nonumber \\
&& \left. ~~~
- 600929280 \zeta_5 g_1^4 g_2^4
+ 389081833 g_1^4 g_2^4
- 102359376 \zeta_3 g_1^3 g_2^5
+ 229171488 g_1^3 g_2^5
+ 39018672 \zeta_3 g_1^2 g_2^6
\right. \nonumber \\
&& \left. ~~~
- 13390272 \zeta_4 g_1^2 g_2^6
- 91542339 g_1^2 g_2^6
+ 419008464 \zeta_3 g_2^8
+ 15023232 \zeta_4 g_2^8
- 598752000 \zeta_5 g_2^8
\right. \nonumber \\
&& \left. ~~~
- 55379079 g_2^8 \right] \frac{1}{11337408} ~+~ \mathcal{O}(g_i^{10})
\end{eqnarray}
where $\zeta_z$ is the Riemann zeta function and $g_i$ in the order symbol
represents either coupling constant. The two (UV) beta-functions are
\begin{eqnarray}
\beta_1(g_i) &=&
- \frac{1}{2} \epsilon g_1
- \left[
- 17 g_1^2
+ 42 g_1 g_2
- 7 g_2^2 \right] \frac{g_1}{36}
+ \left[
- 6617 g_1^4
+ 3591 g_1^3 g_2
- 1988 g_1^2 g_2^2
+ 2625 g_1 g_2^3
- 791 g_2^4 \right] \frac{g_1}{1944}
\nonumber \\
&&
- \left[ 4380480 \zeta_3 g_1^6
+ 2469685 g_1^6
- 1959552 \zeta_3 g_1^5 g_2
+ 9953370 g_1^5 g_2
+ 9897552 \zeta_3 g_1^4 g_2^2
- 3206105 g_1^4 g_2^2
\right. \nonumber \\
&& \left. ~~~
+ 10723104 \zeta_3 g_1^3 g_2^3
- 3644172 g_1^3 g_2^3
+ 2830464 \zeta_3 g_1^2 g_2^4
+ 4790919 g_1^2 g_2^4
- 2231712 \zeta_3 g_1 g_2^5
\right. \nonumber \\
&& \left. ~~~
+ 4461618 g_1 g_2^5
+ 371952 \zeta_3 g_2^6
- 1217531 g_2^6 \right] \frac{g_1}{419904}
\nonumber \\
&&
+ \left[
- 13032226080 \zeta_3 g_1^8
- 571011120 \zeta_4 g_1^8
+ 5888920320 \zeta_5 g_1^8
- 2545129585 g_1^8
- 6517823760 \zeta_3 g_1^7 g_2
\right. \nonumber \\
&& \left. ~~~
+ 2609143488 \zeta_4 g_1^7 g_2
+ 18674530560 \zeta_5 g_1^7 g_2
- 14014594272 g_1^7 g_2
- 41976125856 \zeta_3 g_1^6 g_2^2
- 1654270128 \zeta_4 g_1^6 g_2^2
\right. \nonumber \\
&& \left. ~~~
+ 41686202880 \zeta_5 g_1^6 g_2^2
+ 1287653416 g_1^6 g_2^2
+ 55393105824 \zeta_3 g_1^5 g_2^3
+ 2510431056 \zeta_4 g_1^5 g_2^3
- 53417387520 \zeta_5 g_1^5 g_2^3
\right. \nonumber \\
&& \left. ~~~
- 6038544078 g_1^5 g_2^3
- 62879166000 \zeta_3 g_1^4 g_2^4
- 739159344 \zeta_4 g_1^4 g_2^4
+ 63110638080 \zeta_5 g_1^4 g_2^4
+ 8724733738 g_1^4 g_2^4
\right. \nonumber \\
&& \left. ~~~
+ 26658589104 \zeta_3 g_1^3 g_2^5
- 2891318976 \zeta_4 g_1^3 g_2^5
- 19379969280 \zeta_5 g_1^3 g_2^5
- 8431548300 g_1^3 g_2^5
- 32862267648 \zeta_3 g_1^2 g_2^6
\right. \nonumber \\
&& \left. ~~~
- 1099390320 \zeta_4 g_1^2 g_2^6
+ 30451438080 \zeta_5 g_1^2 g_2^6
+ 3554373284 g_1^2 g_2^6
- 7103920320 \zeta_3 g_1 g_2^7
+ 256456368 \zeta_4 g_1 g_2^7
\right. \nonumber \\
&& \left. ~~~
+ 10777536000 \zeta_5 g_1 g_2^7
- 252392070 g_1 g_2^7
+ 1257025392 \zeta_3 g_2^8
+ 45069696 \zeta_4 g_2^8
- 1796256000 \zeta_5 g_2^8
\right. \nonumber \\
&& \left. ~~~
- 166137237 g_2^8 \right] \frac{g_1}{68024448} ~+~ \mathcal{O}(g_i^{11})
\end{eqnarray}
and
\begin{eqnarray}
\beta_2(g_i) &=&
- \frac{1}{2} \epsilon g_2
- \left[ 6 g_1^3
- 3 g_1^2 g_2
- 13 g_2^3 \right] \frac{1}{12}
+ \left[ 549 g_1^5
- 537 g_1^4 g_2
+ 747 g_1^3 g_2^2
- 420 g_1^2 g_2^3
- 2951 g_2^5 \right] \frac{1}{648}
\nonumber \\
&&
- \left[
- 629856 \zeta_3 g_1^7
+ 1501722 g_1^7
+ 1648512 \zeta_3 g_1^6 g_2
- 469209 g_1^6 g_2
+ 2822688 \zeta_3 g_1^5 g_2^2
- 1281780 g_1^5 g_2^2
\right. \nonumber \\
&& \left. ~~~
+ 1006992 \zeta_3 g_1^4 g_2^3
+ 1066437 g_1^4 g_2^3
- 637632 \zeta_3 g_1^3 g_2^4
+ 1008810 g_1^3 g_2^4
- 563883 g_1^2 g_2^5
+ 2549232 \zeta_3 g_2^7
\right. \nonumber \\
&& \left. ~~~
- 1449209 g_2^7 \right] \frac{1}{139968}
\nonumber \\
&&
+ \left[
- 297429408 \zeta_3 g_1^9
+ 15081552 \zeta_4 g_1^9
+ 876199680 \zeta_5 g_1^9
- 478212348 g_1^9
- 1099291824 \zeta_3 g_1^8 g_2
\right. \nonumber \\
&& \left. ~~~
- 142615728 \zeta_4 g_1^8 g_2
+ 1323630720 \zeta_5 g_1^8 g_2
- 59500523 g_1^8 g_2
+ 3065116464 \zeta_3 g_1^7 g_2^2
- 21205152 \zeta_4 g_1^7 g_2^2
\right. \nonumber \\
&& \left. ~~~
- 3532792320 \zeta_5 g_1^7 g_2^2
+ 488212362 g_1^7 g_2^2
- 6242637600 \zeta_3 g_1^6 g_2^3
+ 289477152 \zeta_4 g_1^6 g_2^3
+ 6083942400 \zeta_5 g_1^6 g_2^3
\right. \nonumber \\
&& \left. ~~~
+ 780950008 g_1^6 g_2^3
+ 4325835456 \zeta_3 g_1^5 g_2^4
+ 182693232 \zeta_4 g_1^5 g_2^4
- 2768567040 \zeta_5 g_1^5 g_2^4
- 2207956584 g_1^5 g_2^4
\right. \nonumber \\
&& \left. ~~~
- 6776964144 \zeta_3 g_1^4 g_2^5
- 38537856 \zeta_4 g_1^4 g_2^5
+ 5538067200 \zeta_5 g_1^4 g_2^5
+ 569390506 g_1^4 g_2^5
- 1272647376 \zeta_3 g_1^3 g_2^6
\right. \nonumber \\
&& \left. ~~~
+ 254531808 \zeta_4 g_1^3 g_2^6
+ 1539648000 \zeta_5 g_1^3 g_2^6
- 261985350 g_1^3 g_2^6
+ 273150144 \zeta_3 g_1^2 g_2^7
- 101570112 \zeta_4 g_1^2 g_2^7
\right. \nonumber \\
&& \left. ~~~
- 62132436 g_1^2 g_2^7
- 4547016000 \zeta_3 g_2^9
- 298260144 \zeta_4 g_2^9
+ 3467318400 \zeta_5 g_2^9
+ 710525229 g_2^9 \right] \frac{1}{7558272}
\nonumber \\
&& +~ \mathcal{O}(g_i^{11}) ~.
\end{eqnarray}
For the renormalization of the masses we define the mixing matrix of anomalous
dimensions by $\gamma_{ij}(g_k)$ where the masses of $\phi_i$ and $z_a$ are
labelled respectively by $1$ and $2$. The presence of the two tags $\mu_i^2$
allows one to extract the four respective renormalization constants by the same
method as \cite{jag1} which results in
\begin{eqnarray}
-\gamma_{11}(g_i) &=&
-\frac{10}{9} g_1^2
+ 25 \left[
- 23 g_1^2
- 21 g_1 g_2
+ 7 g_2^2 \right] \frac{g_1 ^2 }{486}
\nonumber \\
&&
- 35 \left[ 51840 \zeta_3 g_1^4
+ 55841 g_1^4
- 15552 \zeta_3 g_1^3 g_2
- 28374 g_1^3 g_2
+ 5184 \zeta_3 g_1^2 g_2^2
+ 42923 g_1^2 g_2^2
- 2418 g_1 g_2^3
\right. \nonumber \\
&& \left. ~~~~~~~~~
+ 3672 g_2^4 \right] \frac{g_1^2}{104976}
\nonumber \\
&&
+ 5 \left[
- 260546544 \zeta_3 g_1^6
- 263046528 \zeta_4 g_1^6
+ 345487680 \zeta_5 g_1^6
- 1105616335 g_1^6
- 1164844800 \zeta_3 g_1^5 g_2
\right. \nonumber \\
&& \left. ~~~~~
+ 459025056 \zeta_4 g_1^5 g_2
- 88179840 \zeta_5 g_1^5 g_2
- 258947892 g_1^5 g_2
- 1012943232 \zeta_3 g_1^4 g_2^2
- 332797248 \zeta_4 g_1^4 g_2^2
\right. \nonumber \\
&& \left. ~~~~~
+ 1422308160 \zeta_5 g_1^4 g_2^2
- 74040764 g_1^4 g_2^2
- 2119745376 \zeta_3 g_1^3 g_2^3
- 54377568 \zeta_4 g_1^3 g_2^3
+ 1685214720 \zeta_5 g_1^3 g_2^3
\right. \nonumber \\
&& \left. ~~~~~
- 117454092 g_1^3 g_2^3
- 1103962608 \zeta_3 g_1^2 g_2^4
- 55030752 \zeta_4 g_1^2 g_2^4
+ 1913829120 \zeta_5 g_1^2 g_2^4
- 776455043 g_1^2 g_2^4
\right. \nonumber \\
&& \left. ~~~~~
+ 43382304 \zeta_3 g_1 g_2^5
+ 40170816 \zeta_4 g_1 g_2^5
- 119868840 g_1 g_2^5
- 10977120 \zeta_3 g_2^6
- 6695136 \zeta_4 g_2^6
\right. \nonumber \\
&& \left. ~~~~~
+ 68062526 g_2^6 \right] \frac{g_1^2}{34012224} ~+~ \mathcal{O}(g_i^{10})
\end{eqnarray}
\begin{eqnarray}
-\gamma_{21}(g_i) &=&
-g_1^2
+ \left[ 17 g_1^2
- 63 g_1 g_2
- 21 g_2^2 \right] \frac{g_1^2}{18}
\nonumber \\
&&
 - \left[
- 23328 \zeta_3 g_1^4
+ 83771 g_1^4
- 52318 g_1^3 g_2
- 6048 \zeta_3 g_1^2 g_2^2
+ 8813 g_1^2 g_2^2
- 60480 \zeta_3 g_1 g_2^3
+ 7462 g_1 g_2^3
\right. \nonumber \\
&& \left. ~~~
+ 55776 g_2^4 \right] \frac{g_1^2}{3888}
\nonumber \\
&&
+ \left[
- 73183392 \zeta_3 g_1^6
- 7231680 \zeta_4 g_1^6
+ 124688160 \zeta_5 g_1^6
- 34030172 g_1^6
+ 120331008 \zeta_3 g_1^5 g_2
\right. \nonumber \\
&& \left. ~~~
- 10859184 \zeta_4 g_1^5 g_2
- 158124960 \zeta_5 g_1^5 g_2
- 87191811 g_1^5 g_2
- 3123792 \zeta_3 g_1^4 g_2^2
- 12845952 \zeta_4 g_1^4 g_2^2
\right. \nonumber \\
&& \left. ~~~
+ 10886400 \zeta_5 g_1^4 g_2^2
- 47057234 g_1^4 g_2^2
- 14230944 \zeta_3 g_1^3 g_2^3
- 13281408 \zeta_4 g_1^3 g_2^3
- 126554400 \zeta_5 g_1^3 g_2^3
\right. \nonumber \\
&& \left. ~~~
+ 64341480 g_1^3 g_2^3
+ 11941776 \zeta_3 g_1^2 g_2^4
- 18996768 \zeta_4 g_1^2 g_2^4
- 25855200 \zeta_5 g_1^2 g_2^4
- 60530316 g_1^2 g_2^4
\right. \nonumber \\
&& \left. ~~~
- 17720640 \zeta_3 g_1 g_2^5
+ 31652208 \zeta_4 g_1 g_2^5
- 67495680 \zeta_5 g_1 g_2^5
- 58803297 g_1 g_2^5
+ 3592512 \zeta_3 g_2^6
\right. \nonumber \\
&& \left. ~~~
+ 2449440 \zeta_4 g_2^6
- 1753038 g_2^6 \right] \frac{g_1^2}{629856} ~+~ \mathcal{O}(g_i^{10})
\end{eqnarray}
\begin{eqnarray}
-\gamma_{12}(g_i) &=&
 - \frac{5}{3} g_1^2
+ 5 \left[ 2 g_1^2
- 189 g_1 g_2
- 14 g_2^2 \right] \frac{g_1^2}{162}
\nonumber \\
&&
-5 \left[
- 88128 \zeta_3 g_1^4
+ 244250 g_1^4
- 81648 \zeta_3 g_1^3 g_2
- 43974 g_1^3 g_2
+ 45360 \zeta_3 g_1^2 g_2^2
+ 74669 g_1^2 g_2^2
- 27216 \zeta_3 g_1 g_2^3
\right. \nonumber \\
&& \left. ~~~~~
- 139482 g_1 g_2^3
+ 27216 \zeta_3 g_2^4
+ 77273 g_2^4 \right] \frac{g_1^2}{34992}
\nonumber \\
&&
+ 5 \left[
- 209158200 \zeta_3 g_1^6
+ 17128584 \zeta_4 g_1^6
+ 336506400 \zeta_5 g_1^6
- 204713141 g_1^6
+ 309799728 \zeta_3 g_1^5 g_2
\right. \nonumber \\
&& \left. ~~~~~
- 11512368 \zeta_4 g_1^5 g_2
- 430284960 \zeta_5 g_1^5 g_2
- 263394642 g_1^5 g_2
- 27025488 \zeta_3 g_1^4 g_2^2
- 114225552 \zeta_4 g_1^4 g_2^2
\right. \nonumber \\
&& \left. ~~~~~
+ 6531840 \zeta_5 g_1^4 g_2^2
+ 14773990 g_1^4 g_2^2
- 23024736 \zeta_3 g_1^3 g_2^3
- 20575296 \zeta_4 g_1^3 g_2^3
- 507034080 \zeta_5 g_1^3 g_2^3
\right. \nonumber \\
&& \left. ~~~~~
+ 220078110 g_1^3 g_2^3
+ 86388120 \zeta_3 g_1^2 g_2^4
- 31230360 \zeta_4 g_1^2 g_2^4
- 43273440 \zeta_5 g_1^2 g_2^4
- 186362841 g_1^2 g_2^4
\right. \nonumber \\
&& \left. ~~~~~
+ 52390800 \zeta_3 g_1 g_2^5
+ 19350576 \zeta_4 g_1 g_2^5
- 119206080 \zeta_5 g_1 g_2^5
- 166599216 g_1 g_2^5
+ 41504400 \zeta_3 g_2^6
\right. \nonumber \\
&& \left. ~~~~~
- 16819488 \zeta_4 g_2^6
+ 29393280 \zeta_5 g_2^6
- 18765152 g_2^6 \right] \frac{g_1^2}{5668704} ~+~ \mathcal{O}(g_i^{10})
\end{eqnarray}
\begin{eqnarray}
-\gamma_{22}(g_i) &=&
-\left[ - 3 g_1^2
+ 35 g_2^2 \right] \frac{1}{18}
+ \left[ - 633 g_1^4
- 315 g_1^3 g_2
+ 546 g_1^2 g_2^2
+ 959 g_2^4 \right] \frac{1}{486}
\nonumber \\
&&
 -\left[ 1228608 \zeta_3 g_1^6
- 82353 g_1^6
- 1143072 \zeta_3 g_1^5 g_2
- 69300 g_1^5 g_2
+ 2313360 \zeta_3 g_1^4 g_2^2
+ 3746673 g_1^4 g_2^2
\right. \nonumber \\
&& \left. ~~~
+ 489888 \zeta_3 g_1^3 g_2^3
- 2310084 g_1^3 g_2^3
+ 313131 g_1^2 g_2^4
+ 2004912 \zeta_3 g_2^6
+ 9924901 g_2^6 \right] \frac{1}{209952}
\nonumber \\
&&
+ \left[ 186594192 \zeta_3 g_1^8
- 75011184 \zeta_4 g_1^8
- 359134560 \zeta_5 g_1^8
- 323438987 g_1^8
- 724653216 \zeta_3 g_1^7 g_2
\right. \nonumber \\
&& \left. ~~~
+ 531773424 \zeta_4 g_1^7 g_2
- 401708160 \zeta_5 g_1^7 g_2
- 594763890 g_1^7 g_2
- 367325280 \zeta_3 g_1^6 g_2^2
- 614646144 \zeta_4 g_1^6 g_2^2
\right. \nonumber \\
&& \left. ~~~
+ 736464960 \zeta_5 g_1^6 g_2^2
+ 427325500 g_1^6 g_2^2
- 3795924384 \zeta_3 g_1^5 g_2^3
- 194975424 \zeta_4 g_1^5 g_2^3
+ 2478833280 \zeta_5 g_1^5 g_2^3
\right. \nonumber \\
&& \left. ~~~
+ 234595536 g_1^5 g_2^3
- 2130450336 \zeta_3 g_1^4 g_2^4
- 305526816 \zeta_4 g_1^4 g_2^4
+ 4778040960 \zeta_5 g_1^4 g_2^4
- 3088527995 g_1^4 g_2^4
\right. \nonumber \\
&& \left. ~~~
- 328769280 \zeta_3 g_1^3 g_2^5
+ 202078800 \zeta_4 g_1^3 g_2^5
+ 264539520 \zeta_5 g_1^3 g_2^5
- 108887142 g_1^3 g_2^5
+ 351975456 \zeta_3 g_1^2 g_2^6
\right. \nonumber \\
&& \left. ~~~
- 79525152 \zeta_4 g_1^2 g_2^6
+ 546998382 g_1^2 g_2^6
- 3258671472 \zeta_3 g_2^8
- 753366096 \zeta_4 g_2^8
+ 6203342880 \zeta_5 g_2^8
\right. \nonumber \\
&& \left. ~~~
- 1091649300 g_2^8 \right] \frac{1}{11337408} ~+~ \mathcal{O}(g_i^{10})
\end{eqnarray}
\end{widetext}
for $N$~$=$~$3$. In accordance with our definition of $\epsilon=(6-d)/2$, further
definitions of anomalous dimensions, beta-functions, and the mass matrix
in Eqs. (7)-(14) also differ from standard definitions \cite{zinn-justin} by an overall factor of two, to agree with the ones used in \cite{jag1}.
We will provide the standard values of the exponents below.

One aspect of a high loop order computation such as the one undertaken here is
ensuring that it is correct. Given the way we enacted the renormalization
automatically already provides one internal check. This is because the
non-simple poles in $\epsilon$ of the renormalization constants are not
independent. Instead they are precisely determined by all the poles in the
renormalization constant from the lower loop order. We note therefore that this
internal renormalization group consistency check was satisfied in all the
renormalization constants computed which gives us confidence in the correctness
of our calculation. Another hidden check emerges from an observation to do with
the fixed point structure of the $N$~$=$~$3$ case. By solving
$\beta_i(g_j)$~$=$~$0$ for the critical couplings it transpires that there is
a solution where $g^\ast_1$~$=$~$-$~$g^\ast_2$. This was observed originally in
the one-loop calculation \cite{jag13} and as we show in the next section corresponds
to an emergent $U(3)$ symmetry, at which the anomalous dimensions of both fields become identical.
In \cite{jag1} the renormalization group functions for a six-dimensional cubic scalar field theory with $U(N_c)$ symmetry were computed
directly. Specifying those results to $U(3)$ and evaluating the corresponding
critical exponents we find exact agreement with those of (\ref{BaseModel}) at
the emergent $U(3)$ fixed point, which provides a solid check on our $O(N)$
tensor theory. In particular the (standard) anomalous dimensions $\eta_\phi= 2 \gamma_\phi (g_i ^*) $ and $\eta_z = 2 \gamma_z (g_i ^*)$
at the $U(3)$-symmetric fixed point are
\begin{widetext}
\begin{eqnarray}
\eta_\phi ~=~ \eta_z &=& \frac{10}{33} \epsilon ~+~
\frac{1000}{11979} \epsilon^2 ~+~
10 [ 104544 \zeta_3 + 220057 ] \frac{\epsilon^3}{4348377} \nonumber \\
&& +~ 10 [ 5936914368 \zeta_3 + 170772624 \zeta_4 - 5025952800 \zeta_5 - 192710239 ]
\frac{\epsilon^4}{4735382553} ~+~ \mathcal{O}(\epsilon^5)
\end{eqnarray}
where the one loop term is in agreement with \cite{jag13} and the two-loop term is in agreement with \cite{mckane}. The fixed-point values of the
couplings are
\begin{eqnarray}
\left( g_1^\ast \right)^2 ~=~ \left( g_2^\ast \right)^2 &=&
\frac{3}{11} \epsilon ~+~ \frac{1301}{3993} \epsilon^2 ~+~
[ 1672704 \zeta_3 + 3301487 ] \frac{\epsilon^3}{5797836} \nonumber \\
&& +~ [ 107879214960 \zeta_3 + 2732361984 \zeta_4 - 76753591200 \zeta_5 - 6440858957 ]
\frac{\epsilon^4}{9470765106} ~+~ \mathcal{O}(\epsilon^5) ~.~~
\end{eqnarray}
Using these values then the higher order corrections to the universal
exponents corresponding to the two IR relevant directions out of the critical
surface are deduced from the eigenvalues of the mass mixing matrix $\gamma_{ij}$ at the fixed point. We find
\begin{eqnarray}
\theta_1 &=& 2 ~+~ \frac{2}{33} \epsilon ~+~ \frac{6250}{11979} \epsilon^2 ~+~
5 [ 209088 \zeta_3 + 285869 ] \frac{\epsilon^3}{ 1449459 } \nonumber \\
&& +~ 5 [ 3466069200 \zeta_3 + 2049271488 \zeta_4 + 5535604800 \zeta_5 + 5263987741 ]
\frac{\epsilon^4}{ 9470765106 } ~+~ \mathcal{O}(\epsilon^5), \nonumber \\
\theta_2 &=& 2 ~+~ \frac{50}{33} \epsilon ~+~ \frac{3830}{3993} \epsilon^2 ~+~
5 [ 2718144 \zeta_3 + 5472707 ] \frac{\epsilon^3}{ 4348377 } \nonumber \\
&& +~ 5 [ 132436914192 \zeta_3 + 8880176448 \zeta_4 - 117320322240 \zeta_5 + 26569626097 ]
\frac{\epsilon^4}{ 9470765106  } ~+~ \mathcal{O}(\epsilon^5),
\end{eqnarray}
\end{widetext}
where $\theta_i$ are the two eigenvalues of the matrix $2 (\delta_{ij} + \gamma_{ij} (g_i ^*)) $.  Recalling that our $\epsilon= (6-d)/2$ yields
the one-loop term in agreement with \cite{jag13}. The eigenvector corresponding to the larger eigenvalue ($\theta_2$) is also at every order a symmetric combination of the two masses, just as one would expect for the $U(3)$-symmetric theory. This is then yet another check on our calculation.

\section{Emergence of $U(3)$ symmetry}

  Let us next show that on the line $g_1 + g_2 =0$ the last two terms in (1) for $N=3$ can be rewritten in a manifestly
$U(3)$ invariant way. Consider $ Tr (M^3 $)  where $M$ is a three-dimensional, traceless, Hermitean matrix.
It can be expanded in terms of Gell-Mann matrices as
\begin{equation}
M= z_a \Lambda^a + \phi_i S^i,
\end{equation}
where $\Lambda^a$ are the five real, and $S^i$ the three imaginary Gell-Mann matrices, with matrix elements
\begin{equation}
(S^i)_{jk}= i \epsilon_{ijk},
\end{equation}
with $\epsilon_{ijk}$ as the fully antisymmetric tensor. We make no distinction between upper and lower indices, and use them both solely for notational
clarity. Evidently, the three $S^i$ are the generators of $SO(3)$ in the adjoint representation,
and the five $\Lambda^a$ are a second-rank tensorial representation of the same group. Taken together the eight matrices close the algebra  of
$SU(3)$, as well known.

Then,
\begin{eqnarray}
Tr ( M^3 ) &=&
 z_a z_b z_c Tr(\Lambda^a \Lambda^b \Lambda^c) + 3 z_a \phi_i \phi_j Tr (\Lambda^a S^i S^j) \nonumber \\
&& + 3 z_a z_b \phi_k Tr (\Lambda^a \Lambda^b S^k) + \phi_i \phi_j \phi_k Tr ( S^i S^j S^k) ~. \nonumber \\
\end{eqnarray}

 The last two terms are identically zero. Consider first the last term:
 \begin{equation}
 \phi_i \phi_j \phi_k Tr ( S^i S^j S^k) = -i  \phi_i \phi_j \phi_k \epsilon_{inm}\epsilon_{jml} \epsilon_{kln}.
 \end{equation}
 Since $\epsilon_{inm}\epsilon_{jml}= \delta_{il} \delta_{nj}- \delta_{ij} \delta_{ln}$ the last line reduces to
 \begin{equation}
 -i \phi_i \phi_j \phi_k (\epsilon_{kij} - \delta_{ij} \epsilon_{kll}) = 0.
 \end{equation}
The third term in Eq. (20) is proportional to
\begin{equation}
z_a z_b \phi_k Tr (\Lambda^a \Lambda^b S^k)= \frac{i}{2} z_a z_b \phi_k ( \Lambda^a _{ij} \Lambda^b _{jl} + \Lambda^a _{lj} \Lambda^b _{ji} ) \epsilon_{kli},
\end{equation}
where we used the symmetry property of the $\Lambda$ matrices in  the second term. By exchanging the indices $i$ and $l$ in the second term we  can further rewrite this as
\begin{equation}
\frac{i}{2} z_a z_b \phi_k \Lambda^a _{ij} \Lambda^b _{jl} (\epsilon_{kli}+ \epsilon_{kil})=0.
\end{equation}
Finally the second term in Eq. (20) can be simplified as
\begin{equation}
3 z_a \phi_i \phi_j Tr (\Lambda^a S^i S^j)= - 3 z_a \phi_i \phi_j \Lambda^a _{ij},
\end{equation}
using the tracelessness of matrices $\Lambda^a$.

Altogether, the field theory (1) for $N=3$ and when $g_1 = -g_2$ can be written  simply as
\begin{equation}
\mathcal{L} = \frac{1}{4} Tr (\partial_\mu M)^2  + \frac{g_2}{6} Tr (M^3  ),
\end{equation}
which is invariant under a global transformation
\begin{equation}
M \rightarrow U M U^\dagger,
\end{equation}
where $U$ is an arbitrary three-dimensional unitary matrix.

\begin{figure}[t]
\centering
\includegraphics[width=8.5cm]{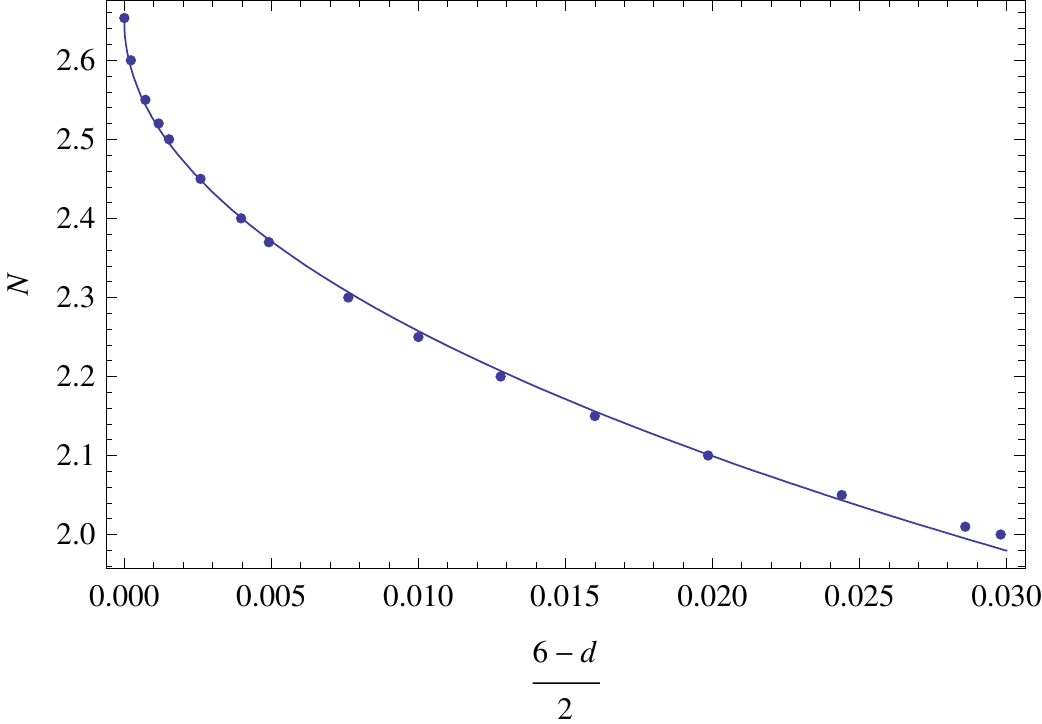}
\caption{ $N_c(\epsilon)$ extracted from three-loop $\beta$-functions, for $2<N<2.6535$. The fit to points is a Taylor expansion in $\epsilon^{1/2}$, given by
Eq. (3) in the introduction. The Wilson-Fisher IR fixed point exists below the curve. }
\label{Fit}
\end{figure}

\section{Evolution of conformal windows}

One-loop beta-functions display IR-stable fixed points when $N$ falls within certain intervals, as discussed in \cite{jag13}. Let us briefly recapitulate these results, as they will be now confirmed and extended. First, besides the Gaussian fixed point at $g_1=g_2=0$, for any $N<4$ there is a fixed point at $g_1=0$ and $g_2\sim \epsilon$ at which the fields $\phi_i$ and $z_a$ decouple, first identified by Priest and Lubensky~\cite{priest}. For $1<N<2.653$ there is an IR-stable fixed point at finite $g_1$ and $g_2$ (the fixed point ``B" in Fig. 2 in \cite{jag13}). When $N\rightarrow 2.653$ the latter fixed point runs away to infinity, in this way terminating the first conformal window within the one-loop  approximation. Increasing $N$ further, at $N=2.999$ two new fixed points at finite $g_1$ and $g_2$ appear, one of which is IR-stable, and the other one is of mixed stability, just like the still present Priest-Lubensky fixed point (Fig. 3 in \cite{jag13}). For $N=3$ the IR-stable fixed point is at the $U(3)$-invariant line $g_1 +g_2 =0 $, at which the fields $\phi_i$ and $z_a$ acquire identical anomalous dimensions. As $N\rightarrow 3.684$ the IR-stable fixed point at finite $g_1$ exchanges stability with the Priest-Lubensky fixed point at $g_1=0$, which now becomes IR-stable. Finally, at $N=4$ the Priest-Lubensky fixed point itself runs away to infinity as well.\cite{priest}

In sum, to the leading order the beta-functions of the theory (1) in tensor representation display two conformal windows: 1) $1<N<2.653$, which surrounds the point $N=2$, and 2) $2.999 < N < 4$, which contains $N=3$ very close to the left edge of the window, and also exhibits the exchange of stability between two fixed points at $ N= 3.684 $.

The values of the boundaries of the conformal windows quoted above may be understood as the leading ($\sim\epsilon^0$) approximation to some unknown functions $N_c (\epsilon)$ \cite{HerbutTesanovic, HerbutBook, giombi}, which may be expressed as Taylor series in $\epsilon^{1/2}$. Tracking the evolution of the fixed-point structure of the two-loop beta functions,  for example, we determine the conformal regions in the $N-\epsilon$ plane, within which an IR-stable $\mathcal{O}(\epsilon)$ fixed point in the theory (1) exists (Figure 1). Adding the third and fourth loop terms turns out to alter this structure only quantitatively, and  only slightly (Figure 3). The principal new feature brought by two loops and higher is the appearance of the $\mathcal{O}(1)$ fixed points already at $d=6$, with mixed stability, which collide with and annihilate the IR-stable $\mathcal{O}(\epsilon)$ fixed points at the upper edges of both conformal regions. This scenario replaces the runaway to infinity of the IR-stable fixed points at $N=2.653$ and $N=4$ observed in the one-loop calculation. In fact, near these boundary values the mixed-stable fixed point in question at $d=6$ becomes weakly coupled, since it is proportional to $\sim(2.653-N)$ for $N<2.653$, and to $\sim(4-N)$ for $N<4$, so that at small epsilon and near and left of both $N=2.653$ and $N=4$ the boundaries of both conformal regions result from a collision of two weakly coupled fixed points. The forms of both lines at $\epsilon=0$ are this way determined solely by the universal coefficients of the two-loop beta-functions. While for the left conformal window the exact form  seems difficult to obtain analytically due to both beta-functions being involved, the right conformal window is determined only by the second beta-function at $g_1=0$, and as $N\rightarrow 4$ is given by
\begin{equation}
\epsilon= \frac{441}{320} (4-N)^2 + \mathcal{O}( (4-N)^3 ).
\end{equation}

The present calculation also agrees well with what was previously found for $N=2$ at two loops \cite{roscher}, which is now a point on the line that goes from $(1, 0.154)$ to $(2.653, 0)$ in the $(N,\epsilon)$ plane. Focusing on this line alone for a moment, we show in Fig. 2 the result of the three-loop calculation, together with a  fit to the points provided by the Eq. (3). The two leading terms in this expansion should be universal.

 One also notices nearly vertical lines on the left edge and inside the right conformal region in Fig. 1. The first corresponds to the collision of the two $\mathcal{O}(\epsilon)$ fixed points as $N$ approaches it from the right, and the second to the  exchange of stability with the Priest-Lubensky fixed point. Both lines are in fact slightly bent towards left for $\epsilon >0$. Finally, we should mention that precisely at $N=3$ with increase of $\epsilon$ two things happen in succession: first, at a very low value of $\epsilon$ the $U(3)$-symmetric IR-stable fixed point and the second mixed-stable $\mathcal{O}(\epsilon)$ fixed point very close to it exchange stability, so that the $U(3)$-symmetric fixed point acquires mixed stability in the $g_1 - g_2$ plane. Increasing $\epsilon$ further at fixed $N=3$ then ultimately brings the new IR-stable (but $U(3)$-asymmetric) fixed point in collision with another $\mathcal{O}(1)$ fixed point, just as at any other $N$. The $U(3)$-symmetric  mixed-stable fixed point is still present at the tip of the right conformal region, and disappears only at a higher value of $\epsilon$, via collision with another IR-unstable fixed point that lies at the $U(3)$-symmetry line.

\begin{figure}[t]
\centering
\includegraphics[width=8.5cm]{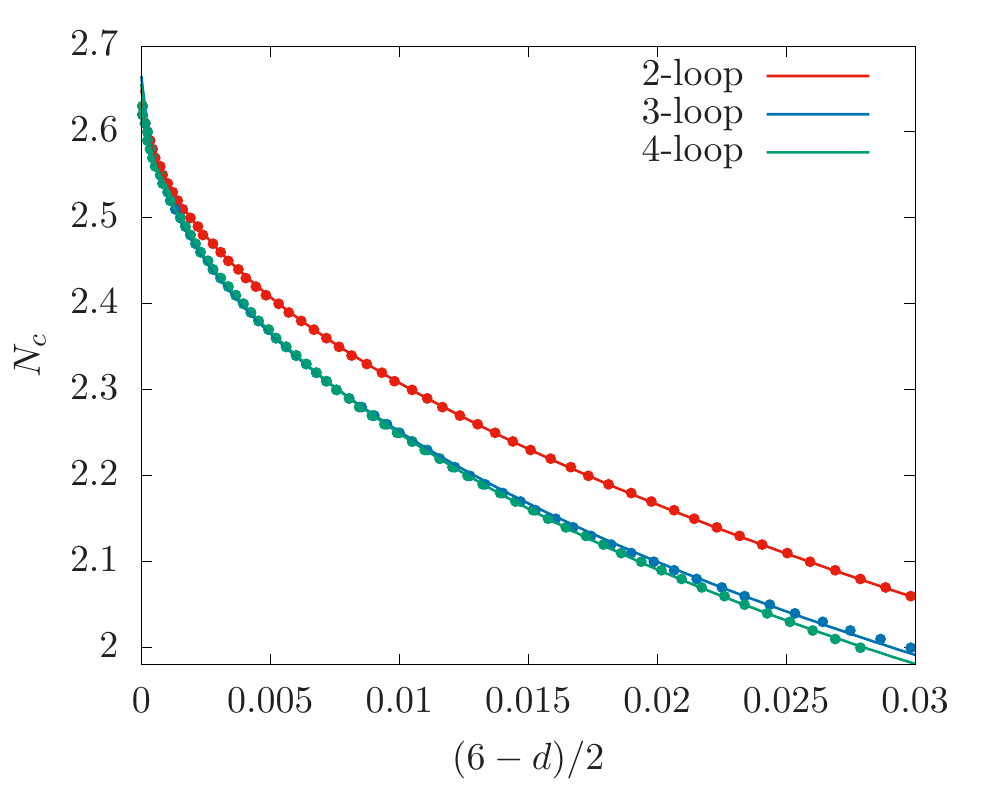}
\caption{ Comparison of $N_c(\epsilon)$ extracted from two-loop, three-loop, and four-loop $\beta$-functions, for $2<N<2.6535$. }
\label{Fit}
\end{figure}

\section{Summary and discussion}

In conclusion, we considered a higher-loop perturbative renormalization group for the tensorial representation of the $O(N)$ $\phi^4$ model
close to six dimensions. In general, the Wilson-Fisher IR-stable $\mathcal{O}(\epsilon)$ fixed point which was previously found in the one-loop calculation with increase of the parameter $\epsilon=(6-d)/2$ becomes complex at some value of $\epsilon$ which is low enough to render the theory trivial in $d=5$. This however, happens only because higher loops in the beta-functions generally introduce a non-trivial $\mathcal{O}(1)$ mixed-stable fixed point which exists even in $d=6$, and which collides with and annihilates  \cite{annihilation} the IR fixed points at some finite $\epsilon$. We nevertheless obtained the critical exponents of the IR fixed points to the order $\epsilon^4$. The $\mathcal{O}(1)$ mixed-stable fixed point becomes weakly coupled in the right corners of both conformal windows, where it can be understood as another example of a Banks-Zaks type of fixed point.

It should be emphasized that the disappearance of the IR fixed points in the theory (1) before the  physical (integer) dimensions such as $d=5$ necessarily comes at a price: since at small $\epsilon$ a real IR fixed point exists both for $N=2$ and $N=3$, it can be moved into the complex plane only through a collision with another fixed point with mixed stability. As a matter of principle, such an additional fixed point may either appear at a finite $\epsilon$, or exist already at $d=6$. We find generally the latter to be true in two- and higher-loop beta-functions for the tensor representation. At $N=2 $ there is no $\mathcal{O}(\epsilon)$ fixed point other than the decoupled Priest-Lubensky fixed point, whereas for $N=3$ there is another non-trivial fixed point with both $g_1\sim \epsilon$ and $g_2 \sim \epsilon $ available for a collision, but the IR-stable fixed point has a larger emergent symmetry. In both cases one of the two $\mathcal{O}(\epsilon)$ fixed points which could in principle collide lie at the RG-invariant higher-symmetry lines: the mixed-stable Priest-Lubensky fixed point at the decoupled line $g_1 =0$ for $N=2$, and the IR-stable fixed point at the $U(3)$-symmetric $g_1 + g_2=0$ line for $N=3$. This feature prevents their annihilation by any fixed point which is not on the same special line, and the only event in which these points can participate is an exchange of stability~\cite{gukov}. Therefore both for $N=2$ and $N=3$ the only option for the ultimate annihilation of the IR-stable  fixed point is by some $\mathcal{O}(1)$ fixed point of mixed stability, which is precisely what we find. The existence of the mixed-stable fixed point may of course be simply an artifact of the perturbative expressions for the $\beta$-functions. If that is the case, however, we hope that the above discussion makes it clear that the theory (1) will have IR fixed points at $d=5$, at least within the usual scheme of $\epsilon$-expansion. If not, the infrared triviality in $d=5$  is a consequence of the existence of a mixed-stable non-trivial fixed point in $d=6$, which is almost equally interesting. The reader should note that the situation in the scalar representation \cite{fei} is radically different: there with a decrease of $N$ the IR fixed point always collides with another $\mathcal{O}(\epsilon)$ fixed point, so they either both exist in say $d=5$ or they both do not. This is similar to what one finds in scalar electrodynamics, for example. \cite{halperin, tesanovic, HerbutBook}

It seems also worth noting that the cases of $N=2$ and of $N=3$ are different in another important respect:
the mixed-stable fixed point that annihilates the IR-stable fixed point at $N=2$ is connected continuously to the perturbative Banks-Zaks-like fixed point near $N=2.653$. Provided the Banks-Zaks fixed point survives the deformation of $N$ from
$N=2.653$ to $N=2$ it may indeed annihilate the Wilson-Fisher fixed point at some finite critical $\epsilon$. This critical value, however, does not necessarily have to be such that $d_c >5$.
We find that critical value of $\epsilon$ at $N=2$, however, to be much smaller than unity, independently of the approximation we take.
At $N=3$, on the other hand, the $\mathcal{O}(1)$ mixed-stable fixed point responsible for the annihilation of the IR-stable fixed point is not connected to the Banks-Zaks fixed point near $N=4$, due to the stability exchange between the Wilson-Fisher and the Priest-Lubensky fixed points around $N=3.684$. Nevertheless, the fixed point that annihilates the IR-stable fixed point at $N<3.684$ eventually also becomes weakly coupled, but at a higher value of $N=4.0057$. As far as we can tell this additional characteristic value of $N$ turns out to be close to $N=4$ purely accidentally.

If we would ignore the observed loss of full IR stability of the $U(3)$-symmetric fixed point with increase of epsilon, it alone would in fact be more robust,  and would become complex due to a collision with an another $U(3)$-symmetric fixed point at a somewhat higher value of $\epsilon=0.21$, in two loops for example. The fixed point that collides with it, however, does not appear to be of Banks-Zaks variety, and its non-perturbative existence may be questioned. Altogether, it is tempting to speculate that the Wilson-Fisher fixed point at $N=3$ has a higher chance of surviving in $d=5$ than its $N=2$ cousin.

 In view of the above uncertainties it would obviously be interesting to apply non-perturbative methods to the model (1), such as recently developed conformal bootstrap~\cite{poland}. Such a study directly in integer dimension has been already performed for the scalar representation~\cite{nakayama}. Extending functional  renormalization group to the tensorial representation (1) is also desirable. Such a calculation was also done previously on the scalar representation \cite{eichorn}, as well as on the tensorial representation, but only for the special case $N=2$~\cite{roscher}. In the latter case the functional renormalization group agreed with the two-loop calculation on the collision of the $\mathcal{O}(\epsilon)$ IR and the $\mathcal{O}(1)$ UV fixed point at a small critical value of $\epsilon$.

\begin{acknowledgments}
The authors are grateful to Igor Boettcher, Lukas Janssen, Michael Scherer, and Omar Zanusso for useful discussions.
J.A.G. gratefully acknowledges the hospitality of the Institute of
Theoretical Physics at Cologne University where this work was initiated as well
as the support of the German Research Foundation (DFG) through a Mercator
Fellowship. I.F.H.  is supported by the NSERC of Canada. D.R. is supported
by the German Research Foundation (DFG) through the Institutional Strategy of the
University of Cologne within the German Excellence Initiative (ZUK 81).

\end{acknowledgments}

\appendix

\begin{widetext}

\section{$O(N)$ renormalization group functions}

In this appendix we record the full expressions for the various renormalization
group functions for the $O(N)$ theory. These are more involved compared with
those of the Hubbard-Stranonovich decomposition of \cite{fei}. When compared to the specific expression for $N=3$ in the main text,
the couplings $g_i$ in the appendix should be understood as $ i g_i$ in the theory (1).

The anomalous dimensions of the two fields are

The large number of terms in comparison with the same anomalous dimensions of
\cite{fei} derive from the group theory defined by (\ref{groupeval}). The two
four loop $\beta$-functions are
%
Finally the entries of the mass mixing matrix are
%
\end{widetext}
The full expressions for all the renormalization group functions are available
in the Supplemental Material, \cite{suppl}.

\end{document}